\documentclass[aps,pra,amsmath,amssymb,showpacs,twocolumn]{revtex4-1}

\usepackage[T1]{fontenc}
\usepackage{amssymb,amsmath}
\usepackage{amsfonts}
\usepackage{hyperref}

\usepackage{mathbbol}
\usepackage{bbm}
\usepackage{dsfont}
\usepackage{xcolor}

\usepackage{graphicx,epsfig,epsf}% Include figure files

\def\C{\mathrm{C}}
\def\T{\mathrm{T}}
\def\A{\mathrm{A}}

\begin{document}

\title{Two-qubit entangling gates between distant atomic qubits in a lattice}
\author{A.~Cesa and J.~Martin}
\affiliation{Institut de Physique Nucl\'eaire, Atomique et de Spectroscopie, CESAM, Universit\'e de Li\`ege, B\^at.\ B15, B - 4000 Li\`ege, Belgium}
\date{June 6, 2017}
\begin{abstract}
Arrays of qubits encoded in the ground-state manifold of neutral atoms trapped in optical (or magnetic) lattices appear to be a promising platform for the realization of a scalable quantum computer. Two-qubit conditional gates between nearest-neighbor qubits in the array can be implemented by exploiting the Rydberg blockade mechanism, as was shown by D.~Jaksch \emph{et al.}\ [Phys.~Rev.~Lett.~\textbf{85}, 2208 (2000)]. However, the energy shift due to dipole-dipole interactions causing the blockade falls off rapidly with the interatomic distance and protocols based on direct Rydberg blockade typically fail to operate between atoms separated by more than one lattice site. In this work, we propose an extension of the protocol of Jaksch \emph{et al.}\ for controlled-Z and controlled-NOT gates which works in the general case where the qubits are not nearest-neighbor in the array. Our proposal relies on the Rydberg excitation hopping along a chain of ancilla non-coding atoms connecting the qubits on which the gate is to be applied. The dependence of the gate fidelity on the number of ancilla atoms, the blockade strength and the decay rates of the Rydberg states is investigated. A comparison between our implementation of distant controlled-NOT gate and one based on a sequence of nearest-neighbor two-qubit gates is also provided.
\end{abstract}

\maketitle

\section{Introduction}

It is now recognized that quantum computing holds the promise of a new technological revolution. For instance, it will enable solving efficiently complex optimization problems or simulating efficiently many-body quantum systems to understand new phases of matter or even biological systems. A wealth of applications in the fields of artificial intelligence and secure communications is also foreseen. The task to build a quantum computer is, however, a considerable one. Different paradigms have been proposed to build a universal quantum computer~\cite{Lad10}, such as cluster-state~\cite{Rau01,Rau03} or gate-based quantum computers~\cite{Nie00}. The latter are composed of a qubit register  on which logic gates are applied. Any unitary operator acting on the register can be approximated with arbitrary accuracy by a sequence of operations from a set of universal quantum gates composed of single-qubit operations and a two-qubit entangling gate~\cite{Nie00}. Although there is always a non-zero probability of error per gate, quantum error correction and fault-tolerant quantum computation open the door to accurate and arbitrarily long quantum computations, provided the error produced by single- and two-qubit gates does not exceed a certain threshold~\cite{Nie00}. High-fidelity quantum gates are thus a major ingredient for scalable quantum computing. Several platforms implementing a universal gate-based quantum computer have been proposed (see e.g.~\cite{Lad10,Neg11} for reviews), which include neutral atoms~\cite{Saf16_1}, photons~\cite{Kok07}, trapped ions~\cite{Sch13,Deb16} and superconducting circuits~\cite{Pla07,Cla08,Dev13}. Cold neutral atoms in optical or magnetic lattices represent a very promising platform due to the long coherence time of the qubits encoded in Zeeman or hyperfine ground states, the possibility to address atoms individually~\cite{Sch04,Lun09,Wei11} and the ability to produce large arrays of qubits~\cite{Neg11,Saf05,Saf10,Saf16_1}. Moreover, deterministic loading of one atom per lattice site in large arrays can be achieved  byrelying on the superfluid-Mott insulator transition in a cloud of ultracold atoms~\cite{Pei03,Wei11}.
 Recently, high fidelity single-qubit gates using microwave fields have been reported in a two-dimensional (2D) array of cesium atoms~\cite{Xia15}. Different schemes implementing two-qubit entangling gates on neutral atoms have been proposed~\cite{Bre99}, one of which relies on the dipole blockade~\cite{Jak00}. By taking advantage of the strong dipole-dipole interactions between atoms in Rydberg states~\cite{Wal08,Gae09,Wil10,Beg13,Bet15}, it is possible to prevent any modifications of the target's atom state conditionally on the control's atom state.
This concept has been demonstrated experimentally with the implementations of two-qubit controlled-NOT (CNOT)~\cite{Zha10,Ise10,Zha12,Mal15} and controlled-phase gates~\cite{Mal15,Mul14}. Note that interactions between atom us in Rydberg states also allows to implement, in principle, quantum gates involving more than two qubits~\cite{Bri07,Ise11}. Most of the protocols for the implementation of two-qubit gates proposed so far operate between atoms that are on adjacent lattice sites. However, in a large array of qubits, it is highly desirable to be able to perform entangling gates between arbitrarily far apart atoms in the lattice. A few proposals addressing this problem have been made~\cite{Wei12,Sod09, Kuz11,Kuz16,Raf12,Raf14}. One idea put forward is to use a spin chain as a quantum bus to perform quantum gates between distant qubits~\cite{Wei12}. It is based on the adiabatic following of the ground state of the spin chain across the paramagnet to crystal phase transition. Another proposal is to use moving carrier atoms of a different species while mediating the quantum gate with molecular states~\cite{Sod09, Kuz11,Kuz16}. Alternatively, it has been suggested to transport the state of the control qubit near the target qubit via optical lattice modulations~\cite{Raf12,Raf14}.

In this work, we propose to use a chain of ancilla noncoding atoms to implement two-qubit entangling gates between atoms arbitrarily far apart in the lattice. The ancilla atoms are used as mediators to connect control and target atoms. Rydberg excitation hopping along the chain of ancilla atoms enables us to modify the state of the target atom conditionally on the state of the control atom via Rydberg blockade. As such, our protocol can be seen as a generalization of the one of Jaksch \emph{et al.}~\cite{Jak00} to the case where the qubits are spatially separated. More specifically, we present protocols that implement either a CNOT gate or a modified control-Z (CZ) gate, represented in the computational basis by the unitary matrices~\cite{footnote4}
\begin{equation}
U_{\mathrm{CNOT}}=\begin{pmatrix}1 & 0 & 0 & 0\\
0 & 1 & 0 & 0\\
0 & 0 & 0 & 1\\
0 & 0 & 1 & 0
\end{pmatrix},
\;\;\tilde{U}_{\mathrm{CZ}}=\begin{pmatrix}1 & 0 & 0 & 0\\
0 & -1 & 0 & 0\\
0 & 0 & -1 & 0\\
0 & 0 & 0 & -1
\end{pmatrix}.
\label{2qubit_gates}
\end{equation}

This paper is organized as follows: In Sec.~\ref{System}, the system and the master equation describing its time evolution are presented. The section also contains a review of the process fidelity used to assess the performance of our protocols in the presence of errors. Sec.~\ref{Protocol} is devoted to the description of the protocol implementing two-qubit entangling gates between atoms arbitrarily far apart in the lattice. In Sec.~\ref{Results}, we present and discuss our results on the effects of dissipation and imperfect blockade on the gate fidelity and compare, in terms of performance, our protocol with implementations using only nearest-neighbor two-qubit gates. Section~\ref{Perspective} discusses some experimental considerations and gives perspectives of our work. A conclusion (Sec.~\ref{Conclusion}) ends this paper.

%%%%%%%%%%%%%%%%%%%%%%%%%%%%%%%%%%%%%%%%%%%%%%%%%%%%%%%%

\section{System and its theoretical description}\label{System}

\subsection{System and Hamiltonian}

\begin{figure}[ht]
\begin{center}
\includegraphics[scale=0.9]{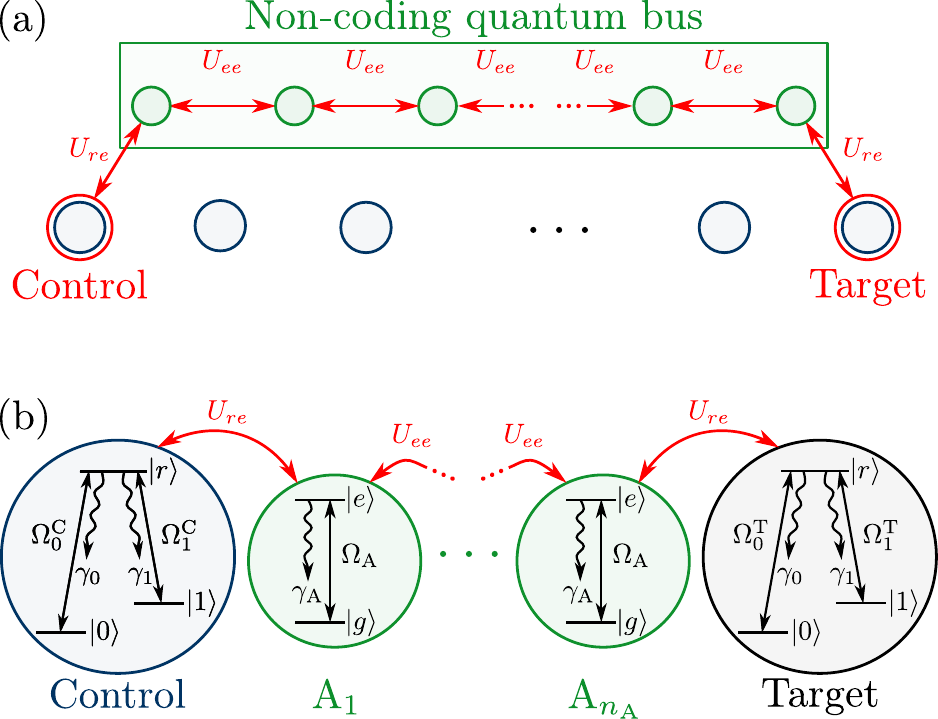}
\caption{(a) Qubits are encoded in a 1D chain of atoms (blue). A parallel chain of ancilla noncoding atoms (green) is used to connect the control and target qubits through the nearest-neighbor Rydberg blockade (red arrows). (b) Internal structure of the qubit and ancilla atoms. For clarity, the subscript denoting the atom to which the different states belong is omitted.}
\label{fig2}
\end{center}
\end{figure}

The physical system that we consider is a one-dimensional (1D) chain of coding atoms (qubit atoms, labeled q) next to a shifted parallel chain of noncoding atoms [ancilla atoms, labeled A; see Fig.~\ref{fig2}(a)]. This system could be implemented e.g.\ by loading an optical lattice with different atomic species~\cite{Bet15}. The protocol that we present in Sec.~\ref{Protocol} also works for 2D or three-dimensional lattices. Here, we consider a 1D lattice merely for computational convenience. The control (C) and target (T) qubits are connected via a chain of $n_{\mathrm{A}}$ ancilla atoms, as illustrated in Fig.~\ref{fig2}(a). 
Each atom, either qubit or ancilla, is assumed to be individually addressable by laser pulses. Qubit atoms are modeled by three-level systems in a $\Lambda$ configuration [see Fig.~\ref{fig2}(b)], where the two lower states $|0_\mathrm{q}\rangle$ and $|1_\mathrm{q}\rangle$ encode the qubit and the upper Rydberg state $|r_\mathrm{q}\rangle$ allows for dipole-dipole interaction with either ancilla or qubit atoms. The transitions $|0_\mathrm{q}\rangle \leftrightarrow |r_\mathrm{q}\rangle$ and $|1_\mathrm{q}\rangle\leftrightarrow |r_\mathrm{q}\rangle$ can be resonantly driven by laser pulses with constant Rabi frequencies $\Omega_0$ and $\Omega_1$ respectively. The Hamiltonian for a single-qubit atom in the basis $\{|0_\mathrm{q}\rangle,|1_\mathrm{q}\rangle,|r_\mathrm{q}\rangle\}$ thus reads
\begin{equation}
\hat{H}_{\mathrm{q}}(t)=\frac{\hbar}{2}\begin{pmatrix}0 & 0 & \Omega_0(t) \\
0 & 0 & \Omega_1(t) \\
\Omega_0^*(t) & \Omega_1^*(t) & 0
\end{pmatrix}.
\end{equation}
As only square pulses will be considered, $\Omega_k(t)=\Omega_k$ ($k=0,1$) whenever a pulse drives the transition $|k_\mathrm{q}\rangle \leftrightarrow |r_\mathrm{q}\rangle$, and $\Omega_k(t)=0$ otherwise.

The ancilla atoms are modelled as two-level systems with a lower-energy state $|g\rangle$ and excited Rydberg state $|e\rangle$ resonantly driven by laser pulses [see Fig.~\ref{fig2}(b)]. In the basis $\{|g\rangle,|e\rangle\}$, the Hamiltonian for a single ancilla atom reads
\begin{equation}
\hat{H}_{\mathrm{A}}(t)=\frac{\hbar}{2}\begin{pmatrix} 0 & \Omega_A(t) \\
\Omega_A^*(t)  & 0
\end{pmatrix},
\end{equation}
with Rabi frequency $\Omega_A(t)=\Omega_A$ whenever a pulse drives the transition $|g\rangle \leftrightarrow |e\rangle$ and $\Omega_A(t)=0$ otherwise. As only the control, target and ancilla atoms take part in the gate protocol, other qubit atoms will not be considered in our description of the two-qubit gate. 

Any two neighboring atoms, either qubit or ancilla, strongly interact via dipole-dipole interactions as soon as one of them is in a Rydberg state. In the strong interaction regime, this leads to an effective energy shift of the doubly excited state, either $|e_ie_{i+1}\rangle$, $|r_\C r_\T\rangle$, or $|e_ir_\mathrm{q}\rangle$ ($\mathrm{q}=\C,\T$) depending on the pair of interacting atoms. When this energy shift is much larger than the atom-laser interaction energy, only one atom can be excited at a time to the Rydberg state and the system is said to exhibit dipole or Rydberg blockade~\cite{Saf10,Wal08,Gae09,Beg13}. Note that this phenomenon can also occur between atoms of different species~\cite{Bet15}. In order to model Rydberg blockade in our system, we add to the system Hamiltonian terms accounting for the energy shifts of the doubly excited states: $U_{rr} |r_{\C} r_{\T}\rangle\langle r_{\C} r_{\T}|$ for the interaction between control and target qubit atoms, $U_{ee} |e_i e_j\rangle\langle e_i e_j|$ with $i \neq j$ for the interaction between ancilla atoms $i$ and $j$, and $U_{re} |r_{\mathrm{q}} e_j\rangle\langle r_\mathrm{q} e_j|$ ($\mathrm{q}=\C,\T$) for the interaction between qubit and ancilla atoms. The total Hamiltonian of our system is thus 
\begin{equation}
\hat{H}(t)=\hat{H}_0(t)+\hat{V}_{\mathrm{dd}},
\end{equation}
with
\begin{equation}
\hat{H}_0(t)=\hat{H}_{\mathrm{q}}^{\C}(t)+\hat{H}_{\mathrm{q}}^{\mathrm{T}}(t)+\sum_{j=1}^{n_{\mathrm{A}}}\hat{H}_{\mathrm{A}}^{j}(t)
\end{equation}
and
\begin{equation}\label{Vdd}
\begin{aligned}
\hat{V}_{\mathrm{dd}}={}&U_{rr} |r_{\C} r_{\T}\rangle\langle r_{\C} r_{\T}|
+\sum_{\mathrm{q}=\C,\T} U_{re} |r_{\mathrm{q}} e_j\rangle\langle r_{\mathrm{q}} e_j|\\
&+\sum_{i\ne j}^{n_{\mathrm{A}}}U_{ee} |e_i e_j\rangle\langle e_i e_j|.
\end{aligned}
\end{equation}
The next-nearest-neighbor energy shifts are also taken into account assuming a resonant dipole-dipole interaction between atoms in Rydberg states~\cite{footnote2}.

%%%%%%%%%%%%%%%%%%%%%%%%%%%%%%%%%%%%%%%%%%%%%%%%%%%%%%%%%%%%%%%%%%%%%%%%%%%%%%%%%%

\subsection{Master equation}

Spontaneous deexcitation of the Rydberg states to lower-energy states is one of the major sources of error in the implementation of quantum gates relying on the dipole blockade mechanism. In our system, we consider that the qubit atoms can decay from the Rydberg state $|r\rangle$ to states $|0\rangle$ and $|1\rangle$ with decay rates $\gamma_0$ and $\gamma_1$, respectively, whereas the ancilla atoms can decay from the Rydberg state $|e\rangle$ to the state $|g\rangle$ with decay rate $\gamma_A$.

In order to take this source of dissipation into account, we solve a master equation for 
the density operator $\hat{\rho}$ describing the global state of the control, target and $n_{\mathrm{A}}$ ancilla atoms. Its standard form reads 
\begin{equation}
\begin{aligned}
\frac{d\hat{\rho}(t)}{dt}= & \frac{1}{i\hbar}\left[\hat{H}(t),\hat{\rho}(t)\right]\\
& +\sum_{\mathrm{q}=\C,\T}\sum_{k=0,1}{\left(\hat{L}_k^{\mathrm{q}}\hat{\rho}(t) (\hat{L}_k^{\mathrm{q}})^\dagger -\frac{1}{2} \left\{(\hat{L}_k^{\mathrm{q}})^\dagger \hat{L}_k^{\mathrm{q}}, \hat{\rho}(t)\right\}\right)}\\
& +\sum_{j=1}^{n_{\A}}\left(\hat{L}_{\A}^j\hat{\rho}(t) (\hat{L}_{\A}^j)^\dagger -\frac{1}{2} \left\{(\hat{L}_{\A}^j)^\dagger \hat{L}_{\A}^j,\hat{\rho}(t)\right\}\right),
\end{aligned}
\label{Linblad}
\end{equation}
with the jump operators $\hat{L}_0=\sqrt{\gamma_0}\,|0\rangle\langle r|$, $\hat{L}_1=\sqrt{\gamma_1}\,|1\rangle\langle r|$, and $\hat{L}_{\A}=\sqrt{\gamma_{\A}}\,|g\rangle\langle e|$. For convenience, we introduce the total decay rate for the qubit atoms, $\gamma_\mathrm{q}=\gamma_0+\gamma_1$.

%%%%%%%%%%%%%%%%%%%%%%%%%%%%%%%%%%%%%%%%%%%%%%%%%%%%%%%%%%%%%%%%%%%%%%%%%%%%%%%%%%

\subsection{Process fidelity}

In order to assess the performance of our protocols implementing CNOT and CZ gates against sources of error, we compute the process fidelity $F_{\mathrm{pro}}$~\cite{Gil05}.
The process fidelity measures the difference between an ideal and real quantum processes. For an ideal unitary quantum process $\hat{U}$, the process fidelity between $\hat{U}$ and the real process specified by its complete positive map $\mathcal{E}(\cdot)$ takes the simple form~\cite{Nie02,Gil05}
\begin{equation}
F_{\mathrm{pro}}(\mathcal{E},\hat{U})=\frac{1}{d^3}\sum_{j=1}^{d^2}{\mathrm{Tr}\left(\hat{U}\hat{A}_j^\dagger \hat{U}^\dagger \mathcal{E}(\hat{A}_j)\right)},
\label{Fprocess}
\end{equation}
where $\{\hat{A}_j:j=1,\ldots,d^2\}$ is a basis for operators acting on a $d$-dimensional Hilbert space that verifies the orthonormalization condition $\mathrm{Tr}(\hat{A}_i^\dagger \hat{A}_j)= d\,\delta_{ij}$. The process fidelity thus corresponds to the overlap between an operator $\hat{A}_j$ evolved with the ideal process and the same operator evolved with  the real process, averaged over all basis operators $\hat{A}_j$. It is related to the average fidelity $F_{\mathrm{av}}$ which quantifies the uniform average over the whole Hilbert space of the overlap between $\hat{U}|\psi\rangle$ and $\mathcal{E}(|\psi\rangle\langle\psi|)$ through~\cite{Nie02,Gil05,Ped07}
\begin{equation}
F_{\mathrm{av}}(\mathcal{E},\hat{U})=\frac{d\,F_{\mathrm{pro}}(\mathcal{E},\hat{U}) +1 }{d+1}.
\label{Faverage}
\end{equation}
The computation of the process fidelity involves the propagation of $d^2$ operators under the process $\mathcal{E}$, which rapidly becomes intractable as $d$ increases. However, lower and upper bounds of the process fidelity can be computed much faster using only two complementary bases of pure states~\cite{Hof05}. Consider a basis of $d$ pure states $\{|\psi_n\rangle:n=1,\ldots,d\}$ and the complementary basis $\{|\phi_k\rangle:k=1,\ldots,d\}$ defined as
\begin{equation}
|\phi_k\rangle=\frac{1}{\sqrt{d}}\sum_{n=1}^d{\exp\left(-i \frac{2 \pi}{d} k n \right)|\psi_n\rangle}.
\end{equation}
Introducing the classical fidelity of the process in a basis $\{|i\rangle:i=1,\ldots,d\}$ of the $d$-dimensional Hilbert space as 
\begin{equation}\label{fidbounds}
F_i(\mathcal{E},\hat{U})=\frac{1}{d}\sum_{i=1}^d{\langle i| \hat{U}^\dagger \mathcal{E}\left(|i\rangle\langle i|\right)\hat{U} |i\rangle},
\end{equation}
the following inequalities hold~\cite{Hof05}
\begin{equation}
F_{\psi_{n}}+F_{\phi_{k}}-1\leq F_{\mathrm{pro}}(\mathcal{E},\hat{U}) \leq \min(F_{\psi_{n}},F_{\phi_{k}}),
\label{HofmannBound}
\end{equation}
where $F_{\psi_{n}}\equiv F_{\psi_{n}}(\mathcal{E},\hat{U})$ and $F_{\phi_{k}}\equiv F_{\phi_{k}}(\mathcal{E},\hat{U})$ are the classical fidelities computed respectively in the bases $\{|\psi_n\rangle\}$ and $\{|\phi_k\rangle\}$. 
The expression $F_{\psi_{n}}+F_{\phi_{k}}-1$ is referred to as the Hofmann bound on process fidelity. For these bounds to be computed, it suffices to propagate $2d$ pure states instead of $d^2$ operators.

As in many other situations considering implementations of quantum computing devices, the system of interest not only contains the qubits but also includes ancilla subsystems or noncoding sublevels needed in order to implement quantum gates. As these ancilla systems or levels are generally in well-defined states before and after the gate operation, computing the fidelity on the whole Hilbert space may lead to overly pessimistic error estimation. In order to avoid this problem, the process fidelity can be computed using only a basis of the relevant qubits subspace~\cite{Ped07}. In the case of our distant-qubit gate protocol, the relevant subspace is spanned by the four states encoding the control and target qubits, whereas all noncoding ancilla atoms are in their ground state. Therefore, the process fidelity will be computed only for this subspace of dimension $d=4$.

%%%%%%%%%%%%%%%%%%%%%%%%%%%%%%%%%%%%%%%%%%%%%%%%%%%%%%%%%%%%%%%%%%%%%%%%%%%%%%%%%%
%%%%%%%%%%%%%%%%%%%%%%%%%%%%%%%%%%%%%%%%%%%%%%%%%%%%%%%%%%%%%%%%%%%%%%%%%%%%%%%%%%
%%%%%%%%%%%%%%%%%%%%%%%%%%%%%%%%%%%%%%%%%%%%%%%%%%%%%%%%%%%%%%%%%%%%%%%%%%%%%%%%%%

\section{Protocol}
\label{Protocol}

Our protocol is a generalization of the one proposed in~\cite{Jak00} for the implementation of a two-qubit quantum gate for the case where the qubits are spatially separated as encountered in arrays of qubits encoded in the internal state of neutral atoms trapped in an optical lattice.
The basic idea of our proposal is to use a chain of ancilla atoms to transfer the Rydberg excitation that the control atom may carry, depending on its initial state, near the target atom. This can again be performed by using the Rydberg blockade. More specifically, we consider the case in which control and target qubits are separated by $n_{\A}$ ancilla non-coding atoms (see Fig.~\ref{fig2}). In the following, we assume that the ancilla atoms are initially prepared in their ground state $|g\rangle$ and that we operate in the strong blockade regime ($U\gg \Omega$). 

%%  Global structure of the pulse sequence (from C to T)

\begin{figure*}[ht]
\centering
\includegraphics[scale=1]{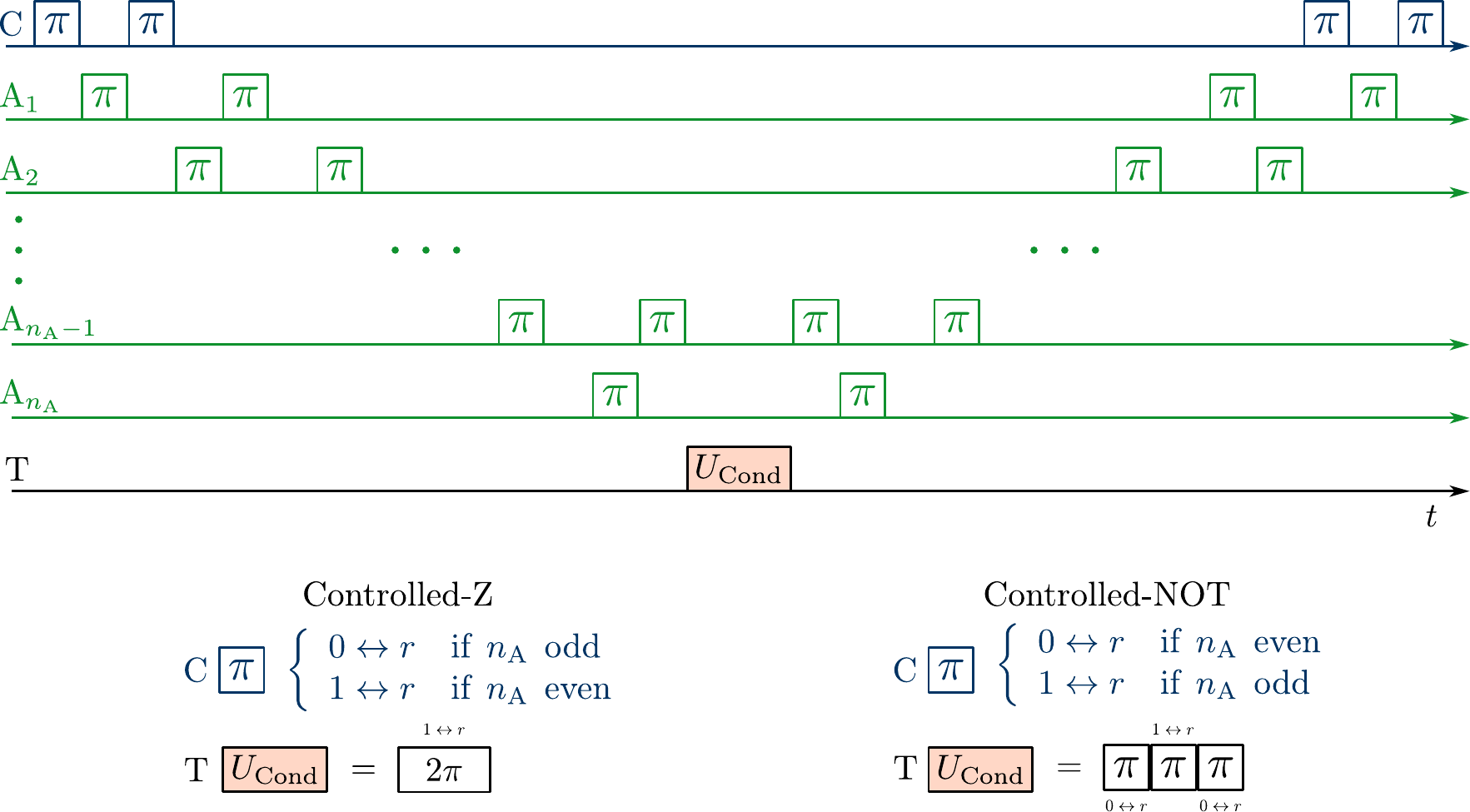}
\caption{
Pulse sequence implementing the CZ and CNOT gates. Note that the pulse sequence on ancilla atoms is the same regardless of the parity of $n_{\A}$.
}
\label{fig3}
\end{figure*}

The generic pulse sequence implementing a given transformation on the target qubit conditionally on the state of the control qubit is illustrated in Fig.~\ref{fig3}.
During the protocol, the transition of the control atom that is to be driven depends both on the length $n_{\A}$ of the chain of ancilla atoms and on the particular two-qubit gate that is to be implemented (in this work either CNOT or modified CZ). The first part of the pulse sequence goes as follows: The first $\pi$ pulse is applied to the control atom and drives only the transition from one of the ground states (either $|0_{\C}\rangle$ or $|1_{\C}\rangle$) to the Rydberg state $|r_{\C} \rangle$. It is followed by a $\pi$ pulse acting on the first ancilla atom $\mathrm{A}_1$. Due to the Rydberg blockade, if the control atom is in $|r_{\C} \rangle$, then $\mathrm{A}_1$ stays in the ground state $|g_1\rangle$, while if the control atom is in the ground-state manifold, then $\mathrm{A}_1$ gets excited to the Rydberg state $|e_1\rangle$. Then, a second $\pi$ pulse is applied on the control atom that brings it back to its initial state.
After these three pulses, the first ancilla atom is in its ground state $|g_1\rangle$ only if the control atom was excited to its Rydberg state $|r_{\C} \rangle$.
Next, two $\pi$ pulses are successively applied to the second and the first ancilla atoms $\mathrm{A}_2$ and $\mathrm{A}_1$. The first one excites $\mathrm{A}_2$ to its Rydberg state $|e_2\rangle$ only if $\mathrm{A}_1$ is in $|g_1\rangle$. The second pulse brings $\mathrm{A}_1$ back to its initial state. Note that if $\mathrm{A}_1$ is initially in $|g_1\rangle$, then the Rydberg blockade due to atom $\mathrm{A}_2$ prevents unwanted excitation of $\mathrm{A}_1$. At this stage of the protocol, the ancilla atom $\mathrm{A}_2$ is in the Rydberg state $|e_2\rangle$ only if the control atom was driven to $|r_{\C} \rangle$ by the very first pulse of the sequence. The same pattern that consists of successive $\pi$-pulses on $\mathrm{A}_{i+1}$ and $\mathrm{A}_{i}$ is applied sequentially to each pair of ancilla atoms, i.e.~for $i=1,\dots,n_\A-1$.

The effect of the pulse sequence above is to produce a hopping of the Rydberg excitation from one atom to the next-nearest-neighbor atom all along the chain separating control and target qubits [see red path in Fig~\ref{fig2}(a)]. More precisely, if the first $\pi$ pulse of the protocol excites the control atom to its Rydberg state $|r_{\C}\rangle$, then the ancilla atoms $\mathrm{A}_{i}$ with even $i$ go through their Rydberg state during the protocol, while those with odd $i$ always stay in the ground state. Conversely, if the control atom is unaffected by the first pulse, i.e.~if it remains in the ground-state manifold, then the ancilla atoms $\mathrm{A}_{i}$ with odd $i$ go through their Rydberg state during the protocol, while those with even $i$ always stay in the ground state.

After this first part of the pulse sequence, a suitable transformation $U_{\textrm{Cond}}$ is applied to the target atom that implements the desired conditional gate. The implementation, which should obviously rely on the dipole blockade mechanism, is shown on the bottom of Fig.~2 for the cases of CZ and CNOT gates. Note that the transition of the control atom to be driven should be chosen in accordance with the parity of the length $n_\A$ of the chain of ancilla atoms. Finally, in order to bring back the ancilla atoms to their initial state, the same pulses as in the first part of the sequence are applied, but this time in reverse order.

During the execution of our protocol, at most one atom at a time is in a Rydberg state, and the Rydberg excitation thus stays localized on a single atom. The number of $\pi$ pulses required in our protocol to implement a long-distance quantum gate is
\begin{equation}
n_{\mathrm{pulse}}=4n_{\mathrm{A}}+2+n_{\T}
\label{n_pulse}
\end{equation}
where $n_{\T}$ is the number of $\pi$-pulses applied on the target atom.
Each $\pi$ pulse leads either to no phase shift when the transition is prevented by the dipole blockade mechanism or to a $\pi/2$ phase shift when the transition is driven resonantly. All atoms, except $\mathrm{A}_{n_{\A}}$ and target atoms, are submitted to four $\pi$ pulses which altogether do not produce any phase shift. Thus, the accumulated phase of the global state comes from only the pulses on $\mathrm{A}_{n_{\A}}$ and on the target atom. If $\mathrm{A}_{n_{\A}}$ is excited to its Rydberg state during the sequence, then it produces a phase shift of $\pi$.

%% Specific case of CZ gate

This generic pulse sequence can be tailored in order to implement either a modified CZ-gate or a CNOT gate [as in Eq.~\eqref{2qubit_gates}]. Let us first consider the case of the modified CZ-gate with an even number of ancilla atoms $n_{\A}=2n$. In that case, the control atom is submitted to (four) pulses driving the transition $|1_{\C}\rangle\leftrightarrow |r_{\C}\rangle$, and the target atom is submitted to a $2\pi$ pulse driving the transition $|1_{\T}\rangle\leftrightarrow |r_{\T}\rangle$. When the control atom is in $|1_{\C}\rangle$, the last ancilla atom $\mathrm{A}_{2n}$ is excited to the Rydberg state $|r_{2n}\rangle$, which produces a $\pi$ phase-shift while the dipole blockade prevents the excitation of the target atom. When the control atom is in $|0_{\C}\rangle$, the last ancilla atom stays in the ground state and the pulse on the target atom produces a $\pi$ phase shift only if the target atom is in $|1_{\T}\rangle$.  Therefore, the only state producing no phase shift is $|0_{\C}0_{\T}\rangle$, and the pulse sequence implements the modified CZ gate~\eqref{2qubit_gates} between two qubits that can be arbitrarily far apart in the lattice. 

For an odd number $n_{\A}=2n+1$ of ancilla atoms, the protocol needs to be slightly amended. In that case, when the control atom is in $|1_{\C}\rangle$, it should not be driven to the Rydberg state by the pulses applied on it, so that the last ancilla atom $\mathrm{A}_{2n+1}$ gets excited to the Rydberg state. Therefore, driving the transition $|0_{\C}\rangle\leftrightarrow |r_{\C}\rangle$ leads to the desired modified CZ gate~\eqref{2qubit_gates}. 

A potential alternative for implementing the modified CZ gate with an odd number of ancilla atoms consists of applying a $\hat{\sigma}_x$ transformation on the control qubit right before and after the protocol shown in Fig.~\ref{fig3}, where the pulses on the control atom drive the transition $|1_{\C}\rangle\leftrightarrow|r_{\C}\rangle$, as was the case for an even number of ancilla atoms. This operation amounts to swapping the role of the states $|0_{\C}\rangle$ and $|1_{\C}\rangle$, which eventually leads to the desired situation where $\mathrm{A}_{2n+1}$ is in the Rydberg state $|e_{2n+1}\rangle$ if the control atom is initially in $|1_{\C}\rangle$ and $\mathrm{A}_{2n+1}$ is in $|g_{2n+1}\rangle$ if the control atom is initially in $|0_{\C}\rangle$. An advantage of this alternative protocol is that, independent of the number of ancilla atoms, only the  $|1_{\C}\rangle\leftrightarrow|r_{\C}\rangle$ transition of the control atoms has to be driven. This simplifies the experimental implementation, at the cost of performing two additional single-qubit gates on the control qubit. 

%% Specific case of CNOT gate

As explained previously, the modified CZ gate can be turned into a CNOT gate using only single-qubit operations~\cite{footnote4}. Nevertheless, it might be useful to directly perform a CNOT gate, which consists of swapping the internal state of the target qubit when the control atom is in $|1_{\C}\rangle$. This can be achieved by applying three $\pi$ pulses on the target atom driving successively the transitions $|0_{\T}\rangle\leftrightarrow |r_{\T}\rangle$, $|1_{\T}\rangle\leftrightarrow |r_{\T}\rangle$, and $|0_{\T}\rangle\leftrightarrow |r_{\T}\rangle$. In the absence of Rydberg excitation near the target atom that would induce Rydberg blockade, this sequence of three pulses swaps the coding state of the target atom. Note that regardless of the state of the target atom, only two pulses out of three effectively affect the target atom. Therefore, this operation on the target atom leads to a $\pi$ phase shift. In order to ensure that the swap operation is performed only if the control atom is in $|1_{\C}\rangle$, the transition $|0_{\C}\rangle\leftrightarrow|r_{\C}\rangle$ ($|1_{\C}\rangle\leftrightarrow|r_{\C}\rangle$) must be driven on the control atom if the number of ancilla atoms is even (odd). The resulting protocol implements a CNOT gate up to a global phase factor of $-1$. 

%%%%%%%%%%%%%%%%%%%%%%%%%%%%%%%%%%%%%%%%%%%%%%%%%%%%%%%%

\section{Results and discussion}
\label{Results}

In this section, we discuss the results of our simulations based on the resolution of the master equation~\eqref{Linblad} for the pulse sequences presented above. For small numbers of ancilla atoms ($n_{\mathrm{A}}\leqslant 5$) the master equation is directly solved for $\hat{\rho}(t)$, while for larger numbers of ancilla atoms ($5<n_{\mathrm{A}}\leqslant 9$) it is solved using a Monte Carlo wave-function approach~\cite{Dal92,Dum92,Mol93,Ple98,Joh12}. In the former case, the exact process fidelity \eqref{Fprocess} is computed, while in the latter case only lower and upper bounds given in Eq.~\eqref{fidbounds} are evaluated. For an odd number of ancilla atoms, we performed the simulations for the alternative protocol by relying on a swap of the internal states of the control qubit. In our simulations, we consider that all transitions are driven with identical Rabi frequencies, i.e.,~$\Omega_0=\Omega_1=\Omega_A=\Omega>0$, which sets a natural frequency unit. We consider square pulses without any delay time between two consecutive pulses. We also take identical energy shifts of the doubly excited states between nearest-neighbor atoms, i.e.,~$U_{rr}=U_{re}=U_{ee}=U$ with $U\gg \Omega$. In all recent experiments demonstrating two-qubit gates based on Rydberg blockade~\cite{Zha10,Ise10,Zha12, Mal15,Mul14}, the Rabi frequency is of the order of $1$~MHz. Moreover, Rydberg states with principal quantum number $n\approx 80$ have a lifetime of the order of 1~ms at cryogenic temperature~\cite{Bet09}.  Accordingly, we choose the decay rates $\gamma_\mathrm{q}=\gamma_0+\gamma_1$ and $\gamma_\mathrm{A}$ to vary with $\gamma_i/\Omega$ ranging from $0$ to $0.01$ ($i=\mathrm{A},\mathrm{q}$).

\subsection{Gate fidelity with respect to the dipole blockade shift}

In this section, we discuss the effects of imperfect blockade on the performance of our protocols. The process fidelity of both gates was numerically computed for values of the dipole blockade shift $U/\Omega$ ranging from $1$ to $200$ and for up to five ancilla atoms in the absence of dissipation ($\gamma_i=0$, $i=0,1,\A$). In this situation, the gate error originates from imperfect blockade. In the strong-blockade regime ($U\gg\Omega$), the probability of double excitation to the Rydberg states is proportional at leading order to $P_{2}\propto\Omega^2/U^2$~\cite{Saf05,Wal08}. In this regime, we expect the gate error also to be proportional to $P_{2}$, which is indeed confirmed by our numerical simulations (data not presented). We observe that the ratio of the gate error $1-F_{\mathrm{proc}}$ to the probability of double excitation $P_{2}$ is constant for $U/\Omega\gtrsim 25$ for both gates, regardless of the number of ancilla atoms.
Thus, in the regime of strong blockade without any dissipation, the process fidelity can be accurately approximated by
\begin{equation}
F_{\mathrm{pro}}^{\gamma_i=0}\left(\frac{U}{\Omega}\right)=1-\alpha \left(\frac{U}{\Omega}\right)^{-2},
\label{F_U}
\end{equation}
with $0.1\lesssim\alpha\lesssim 2$ being a constant whose value depends only on $n_{\mathrm{A}}$ and $U_{\mathrm{Cond}}$. More precisely, $\alpha$ depends only on the parity of the number $n_{\mathrm{A}}$ of ancilla atoms. For the CZ gate, $\alpha\approx 0.5$ for $n_{\mathrm{A}}=0$ and odd $n_{\mathrm{A}}$, while $\alpha\approx 1.7$ for even $n_{\mathrm{A}}$. In the case of the CNOT gate, $\alpha\approx 0.4$ for $n_{\mathrm{A}}=0$, $\alpha\approx 0.1$ for odd $n_{\mathrm{A}}$ and $\alpha\approx 2$ for even $n_{\mathrm{A}}$.
We attribute the differences in $\alpha$ to the dependence on the parity of $n_{\mathrm{A}}$ of the way errors arising from double Rydberg excitation propagate along the chain of ancilla atoms. 

%%%%%%%%%%%%%%%%%%%%%%%%%%%%%%%%%%%%%%%%%%%%%%%%%%%%%%%%%%%%%%%%%%%%%%%%%%%%%%%%%%

\subsection{Gate fidelity with respect to the dissipation rate}

%%%%%%%%%%%%%%%%%%CZ dissipation%%%%%%%%%%%%%%%%%%%%
We now turn to a discussion of the effects of dissipation on the process fidelity. For this purpose, the doubly excited state energy shift is set to $U_{rr}/\Omega=U_{re}/\Omega=U_{ee}/\Omega=200$. At this value of $U/\Omega$ and in the absence of dissipation, the gate error $1-F_{\mathrm{proc}}$ is smaller than $10^{-4}$ for every number of ancilla atoms we will consider.

\begin{figure}[!ht]
\centering
\includegraphics[scale=0.975]{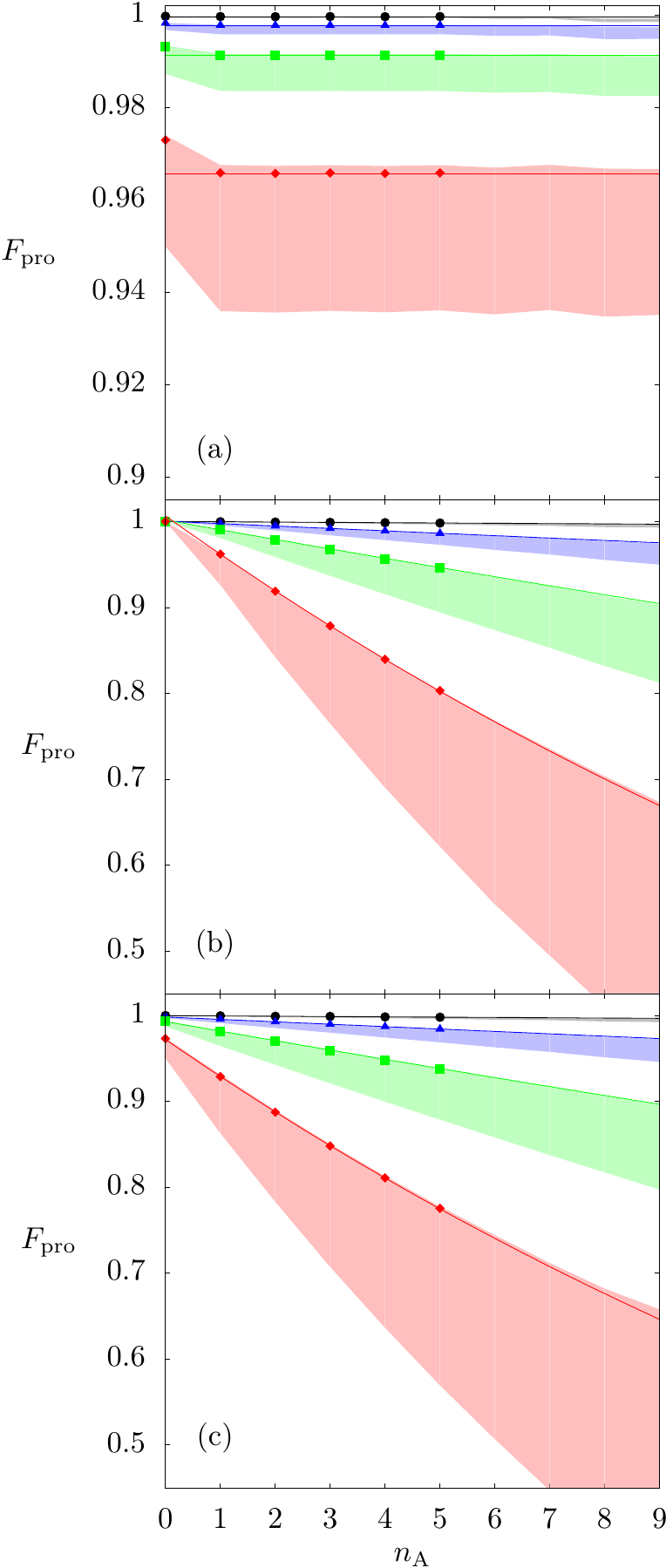}
\caption{Process fidelity $F_{\mathrm{pro}}$ of the distant-qubit protocol implementing a modified CZ gate as a function of $n_{\mathrm{A}}$ for different values of the decay rates: from top to bottom $\gamma/\Omega=4\times 10^{-5}$ (black, dot), $32\times 10^{-5}$ (blue, triangle), $128\times 10^{-5}$ (green, square), and  $512\times 10^{-5}$ (red, diamond). (a) $\gamma_0=\gamma_1=\gamma/2$ and $\gamma_\mathrm{A}=0$, (b) $\gamma_0=\gamma_1=0$ and $\gamma_\mathrm{A}=\gamma$, and (c)  $\gamma_0=\gamma_1=\gamma/2$, $\gamma_\mathrm{A}=\gamma$. The lower and upper bounds on the process fidelity~\eqref{HofmannBound} delimit the shaded area. The symbols are the values of $F_{\mathrm{pro}}$ obtained through numerical simulations and the lines correspond to the fits of the data using~\eqref{Fpro_CZ} with $t_\mathrm{q}$ and $t_\mathrm{A}$ given by~\eqref{t_Q_CZ} and~\eqref{t_A_CZ} with $t_{\mathrm{eff}}^\pi$ as only a fitting parameter (see text). }
\label{fig11}
\end{figure}

\paragraph{Modified CZ gate} The results of our simulations for the process fidelity of the modified CZ gate are displayed in Fig.~\ref{fig11}. Figure~\ref{fig11}(a) and~\ref{fig11}(b) show the process fidelity (dots) in the case of no dissipation on the ancilla and qubit atoms, respectively. Figure~\ref{fig11}(c) shows the process fidelity for identical decay rates for qubit and ancilla atoms. The upper and lower bounds for $F_{\mathrm{pro}}$ [see Eq.~\eqref{HofmannBound}] delimit the shaded areas. Our results show that the actual process fidelity is consistently very close to the upper bound.
For $\gamma_{\mathrm{A}}=0$ and $\gamma_0=\gamma_1=\gamma/2>0$ [dissipation only on the qubit atoms, Fig~\ref{fig11}(a)], the process fidelity no longer depends on the number of ancilla atoms. This is an immediate consequence of our protocol in which the accumulated time spent by the qubit atoms in their Rydberg state does not depend on $n_{\mathrm{A}}$. There is, however, one exception when there are no ancilla atoms ($n_{\mathrm{A}}=0$) because in this case only two $\pi$ pulses are applied to the control atom instead of four, which reduces errors caused by the decay of the control atom from the Rydberg state to the ground-state manifold. 
For $\gamma_{\mathrm{A}}=\gamma$ and $\gamma_0=\gamma_1=0$ [dissipation only on the ancilla atoms, middle panel Fig.~\ref{fig11}(b)], the process fidelity decreases with $n_{\mathrm{A}}$.
For identical decay rates on the qubit and ancilla atoms [Fig.~\ref{fig11}(c)], the process fidelity displays the combined features of the two previous cases. For $n_{\mathrm{A}}=0$, it starts at the value for $\gamma_{\mathrm{A}}=0$ and decreases with $n_{\mathrm{A}}$ as in the case where dissipation acts only on the ancilla atoms [Fig.~\ref{fig11}(b)]. In all cases, the process fidelity decreases with the total decay rate $\gamma$ of the qubit atoms. 

In the protocol depicted in Fig.~\ref{fig3}, dissipation occurs either in the interval between two $\pi$ pulses when an atom, either a qubit or an ancilla, is in its Rydberg state or during one of the pulses. In the former case, the probability for an atom to stay in its Rydberg state decays as $\exp(-\gamma t)$, with $\gamma$ being the decay rate and $t$ being the time since the atom is in its Rydberg state. In the latter case, the probability of unwanted deexcitation from the Rydberg state during a pulse has to be evaluated from the exact solution of the master equation for a decaying two-level atom. 
The probability to be in the target state after a $\pi$ pulse can be written in the form $\exp(-\gamma t_{\mathrm{eff}}^\pi)$, where $t_{\mathrm{eff}}^\pi$ is interpreted as the effective time spent by the atom in the decaying Rydberg state during the pulse. If the decay rates are low enough ($\gamma/\Omega\ll 1$),  $t_{\mathrm{eff}}^\pi$ is constant up to corrections of order $\gamma/\Omega$, for both exciting and deexciting $\pi$ pulses. By equating $\exp(-\gamma t_{\mathrm{eff}}^\pi)$ with the probability for the atom to be in the target state after the $\pi$ pulse as evaluated from the exact solution of the master equation, we obtain $\Omega t_{\mathrm{eff}}^\pi/\pi=3/8$. because in our protocol, only one atom can be in an excited state at a time, the effects of dissipation on the different atoms simply add up, and the process fidelity is determined by the cumulated time spent by the atoms in the decaying Rydberg states during the execution of the protocols.
If the dissipation rates are low enough to ensure that there is at most one decay (quantum jump) during the whole protocol, the process fidelity can be approximated by
\begin{equation}
F_{\mathrm{pro}}\left(\frac{U}{\Omega},\gamma_\mathrm{q},\gamma_{\mathrm{A}}\right)\approx F^{\gamma=0}_{\mathrm{pro}}\left(\frac{U}{\Omega}\right)e^{- \gamma_\mathrm{q}t_\mathrm{q}}e^{- \gamma_{\mathrm{A}} t_\mathrm{A}(n_{\mathrm{A}})}.
\label{Fpro_CZ}
\end{equation}
In Eq.~(\ref{Fpro_CZ}), $F^{\gamma=0}_{\mathrm{pro}}\left(U/\Omega\right)$, given by Eq.~\eqref{F_U}, takes into account the effects of imperfect blockade, $\gamma_\mathrm{q}$ is the total decay rate of the qubit atoms, $t_\mathrm{q}$ is the effective cumulated time spent by the qubit atoms in the Rydberg states averaged over all possible qubit initial states and $t_\mathrm{A}$ is the effective total time spent by the ancilla atoms in the Rydberg states.
Both times $t_\mathrm{q}$ and $t_\mathrm{A}$ can be directly evaluated for the pulse sequence depicted in Fig.~\ref{fig3}. A total of six $\pi$ pulses are applied to the qubit atoms (control and target). Depending on the control atom's initial state, either the four pulses on the control atom or the $2\pi$ pulse on the target atom lead to an excitation to the Rydberg state, but not both at the same time as a consequence of the dipole blockade. Depending on its initial state, the control atom either spends the duration of two $\pi$ pulses in the Rydberg state or stays in the ground state.
Thus, by averaging over all possible initial states, we obtain
\begin{equation}
\Omega t_\mathrm{q}=\frac{2\pi+6\,\Omega t_{\mathrm{eff}}^\pi}{2}.
\label{t_Q_CZ}
\end{equation}
As for the ancilla atoms, they are each submitted to four $\pi$ pulses except for the last one which is submitted to only two $\pi$ pulses. Only half of these pulses bring the ancilla atoms to their Rydberg state, in which they spend in total the duration of four $\pi$ pulses.
This leads eventually to
\begin{equation}
\Omega t_\mathrm{A}(n_{\mathrm{A}})=\frac{4\pi n_{\mathrm{A}}+(4 n_{\mathrm{A}}-2)\Omega t_{\mathrm{eff}}^\pi}{2}.
\label{t_A_CZ}
\end{equation}
Our data for the process fidelity display excellent agreement with Eq.~\eqref{Fpro_CZ}. In fact, by fitting our data by Eq.~\eqref{Fpro_CZ} with $t_{\mathrm{eff}}^\pi$ as the only parameter, we get $\Omega t_{\mathrm{eff}}^\pi/\pi\approx 0.40$, in good agreement with our previous estimate of $3/8$. The fits are shown by solid lines in Fig.~\ref{fig3}. The upper and lower bounds on the fidelity~\eqref{HofmannBound} follow similar behavior with respect to the dissipation rate and the number of ancilla atoms. 

%%%%%%%%%%%%%%%%%%CNOT dissipation%%%%%%%%%%%%%%%%%%%%

\begin{figure}[!ht]
\centering
\includegraphics[scale=0.975]{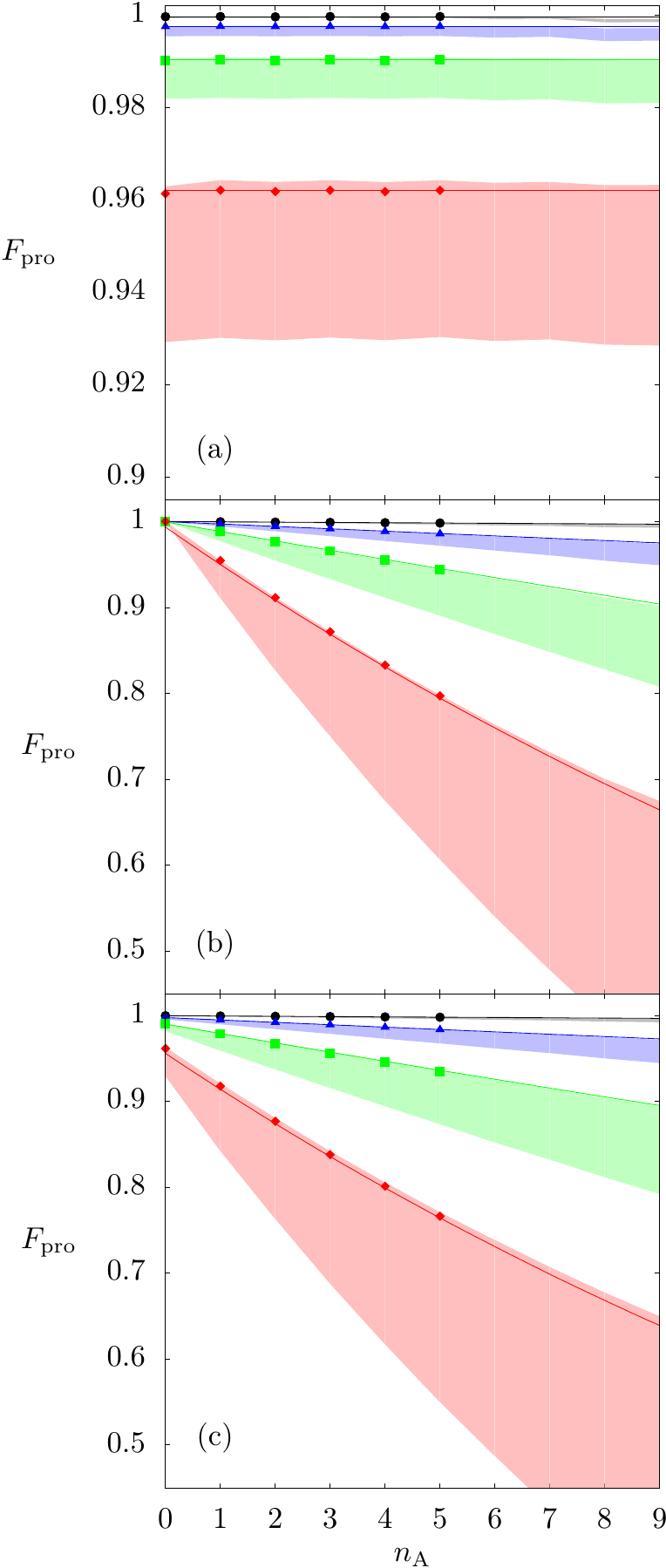}
\caption{Same as in Fig.~\ref{fig11} but for the CNOT gate. The data for $F_{\mathrm{pro}}$ have been fitted using Eq.~\eqref{Fpro_CZ} with $t_\mathrm{q}$ and $t_\mathrm{A}$ given by~\eqref{t_Q_CNOT} and~\eqref{t_A_CNOT} with $t_{\mathrm{eff}}^\pi$ as only a fitting parameter (see text).}
\label{fig12}
\end{figure}

\paragraph{CNOT gate} The results of our simulations for the process fidelity of the CNOT gate are displayed in Fig.~\ref{fig12}. Like in Fig.~\ref{fig11}, Fig.~\ref{fig12}(a) and Fig.~\ref{fig12}(b) show the process fidelity (dots) in the case of no dissipation on the ancilla and qubit atoms respectively. Figure~\ref{fig12}(c) shows the process fidelity for identical decay rates for qubit and ancilla atoms.
Overall, the process fidelity of the CNOT gate behaves, as a function of the decay rate and the number of ancilla atoms, in a  way similar to the modified CZ gate. In the absence of dissipation acting on the ancilla atoms [Fig.~\ref{fig12}(a)], the process fidelity does not depend on $n_{\mathrm{A}}$. When the dissipation acts only on the ancilla atoms [Fig.~\ref{fig12}(b)], the fidelity decreases with $n_{\mathrm{A}}$.  For identical decay rates for the qubit and ancilla atoms [Fig.~\ref{fig12}(c)], the process fidelity displays the combined features of the two previous cases. 
Like for the modified CZ gate, dissipation acts only when the atoms are evolving freely in their Rydberg state or during the pulses that drive qubit or ancilla atoms into their Rydberg states, and thus, the process fidelity of the CNOT gate is determined by the cumulated time spent by the atoms in the Rydberg states.
A  counting argument similar to that for the  modified CZ gate can be made, which leads to an approximation of the form of Eq.~\eqref{Fpro_CZ} for the process fidelity with
\begin{equation}
\Omega t_\mathrm{q}=\frac{2\pi+7\, \Omega t_{\mathrm{eff}}^\pi}{2}
\label{t_Q_CNOT}
\end{equation}
and
\begin{equation}
\Omega t_\mathrm{A}(n_{\mathrm{A}})=\frac{4\pi n_{\mathrm{A}} + \pi +(4 n_{\mathrm{A}}-2)\Omega t_{\mathrm{eff}}^\pi}{2}.
\label{t_A_CNOT}
\end{equation}
In Eq.~\eqref{t_A_CNOT}, the last two  terms in the numerator account for the facts that the last ancilla atom may spend the duration of five $\pi$ pulses in its Rydberg state instead of four and that only two  $\pi$ pulses are applied on it, respectively.
Again, our data for the process fidelity display excellent agreement with Eq.~\eqref{Fpro_CZ}. In fact, by fitting our data by Eq.~\eqref{Fpro_CZ} with $t_{\mathrm{eff}}^\pi$ as the only parameter, we get $\Omega t_{\mathrm{eff}}^\pi/\pi\approx 0.39$. This value is similar to the one obtained for the case of the CZ gate. The results of this fit are illustrated on Fig.~\ref{fig12}. 
%%%%%%%%%%%%%%%%%%%%%%%%%%%%%%%%%%%%%%%%%%%%%%%%%%%%%%%%%%%%%%%%%%%%%%%%%%%%%%%%%%

\subsection{Comparison with a sequence of nearest-neighbor CNOT gates}\label{Comparison}

It is interesting to compare the fidelity of our protocol for the distant-qubit CNOT gate with an implementation relying on a sequence of nearest-neighbor two-qubit gates.
In the geometrical configuration illustrated in Fig.~\ref{fig2} where the two chains of atoms are displaced, performing our distant-qubit protocol for $n_{\mathrm{A}}$ ancilla atoms corresponds to the control and target qubits being separated by $n_{\mathrm{A}}-1$ other qubits. An obvious advantage of our protocol is that only the two qubits involved in the gate are manipulated and thus prone to errors (we recall that ancilla atoms are non coding). In contrast, for a sequence of nearest-neighbor CNOT gates, all the qubits in between the control and target qubits are submitted to quantum gates and thus potentially prone to errors.

\begin{figure}[ht]
\begin{center}
\includegraphics[scale=1]{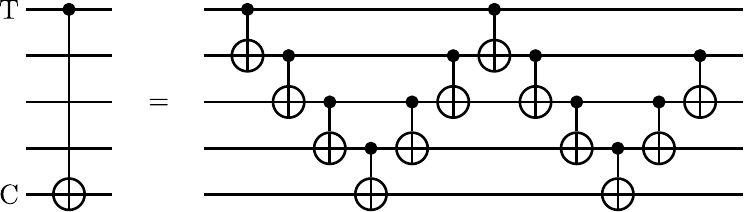}
\caption{Possible implementation of a CNOT quantum gate between non adjacent qubits using only nearest-neighbor CNOT gates~\cite{Rah15}. 
}
\label{fig13}
\end{center}
\end{figure}

The number of pulses needed to perform a distant-qubit CNOT gate with $n_{\mathrm{A}}$ ancilla atoms, given in Eq.~\eqref{n_pulse}, is $4n_{\mathrm{A}}+2+n_{\T}$. The same operation can be performed with $4(n_{\mathrm{A}}-1)$ nearest-neighbor CNOT gates~\cite{Sae11,Rah15}, which amounts to applying $20(n_{\mathrm{A}}-1)$ pulses to the register of qubits, as illustrated in Fig.~\ref{fig13} in the case $n_{\mathrm{A}}-1=3$~\cite{footnote3}.

\begin{figure}[ht]
\begin{center}
\includegraphics[scale=0.9]{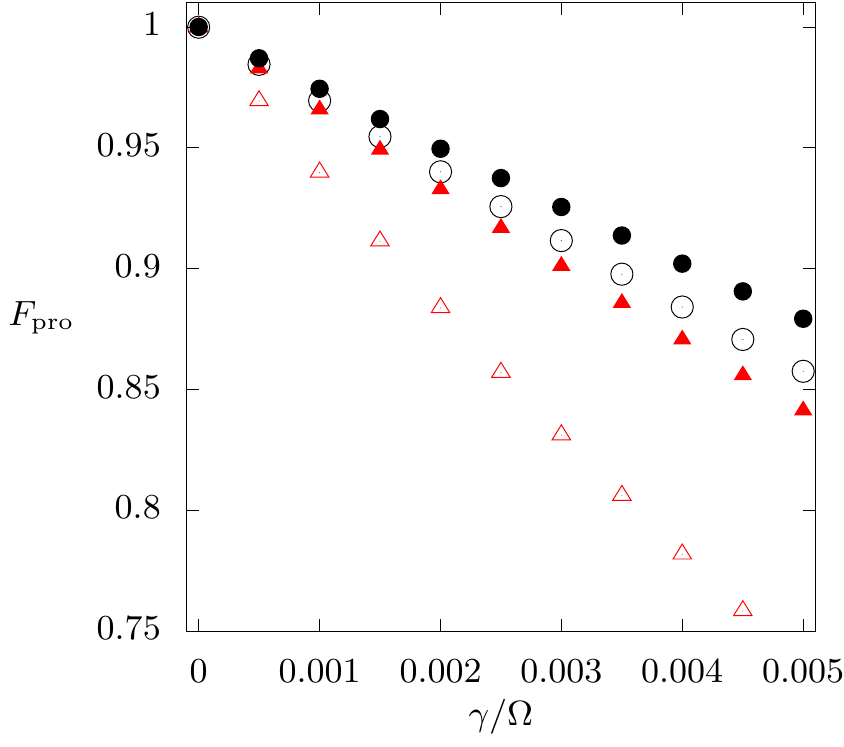}
\caption{Comparison of the process fidelity obtained using only the nearest neighbour CNOT gate~\cite{Rah15} and the distant-qubit protocol using non coding ancilla atoms with $\gamma_\mathrm{A}=\gamma_\mathrm{q}=\gamma$ and $\gamma_0=\gamma_1$. The black dots correspond to the case of two ancilla atoms while the black open circles represent the corresponding fidelity for the nearest-neighbor-qubit implementation with a single qubit between the control and target atoms. The red solid triangles represent the process fidelity in the case of three ancilla atoms and the red open triangles correspond to the case where the control and target are separated by two other qubits.}
\label{fig14}
\end{center}
\end{figure}

\begin{figure}[ht]
\begin{center}
\includegraphics[scale=1]{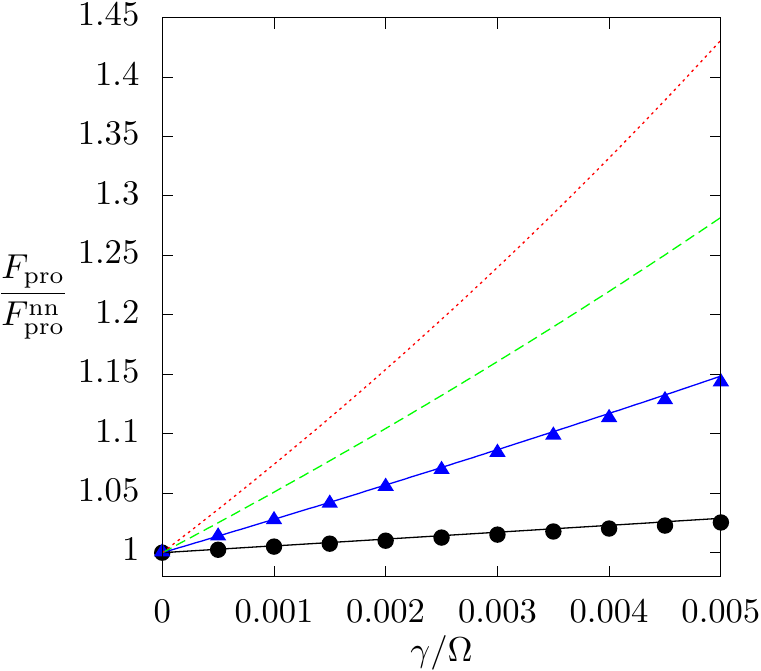}
\caption{Gain in the fidelity~\eqref{gain_nn} resulting from the use of non coding ancilla atoms in a distant-qubit CNOT gate as a function of the decay rate ($\gamma_\mathrm{A}=\gamma_\mathrm{q}=\gamma$ and $\gamma_0=\gamma_1$) for (from bottom to top) $n_{\mathrm{A}}=2$ (black, dot), $3$ (blue, triangle), $4$ (green, dashed), and $5$ (red, dotted). The symbols correspond to the gain computed from our numerical simulations.}
\label{fig15}
\end{center}
\end{figure}

Even for next-nearest-neighbor qubits ($n_{\mathrm{A}}=2$), our protocol requires a smaller number of pulses ($13$ pulses instead of $20$), resulting in a higher process fidelity. This is exemplified in Fig.~\ref{fig14}, where we compare the process fidelity~\eqref{Fprocess} of our protocol with the one based on a sequence of nearest-neighbor CNOT gates~\cite{footnote1}. The process fidelity is plotted as a function of the decay rate when control and target qubits are separated by two and three qubits, respectively.
Our protocol always leads to a higher fidelity. A simple estimate of the gain in fidelity can be made in the case of low dissipation rates following a reasoning similar to that in the previous sections.
For strong blockade and identical decay rates on qubit and ancilla atoms ($\gamma_\mathrm{A}=\gamma_\mathrm{q}=\gamma$ and $\gamma_0=\gamma_1$), the ratio of process fidelities is approximately given by
\begin{equation}
\frac{F_{\mathrm{pro}}}{F_{\mathrm{pro}}^\mathrm{nn}}\approx\exp \left( \frac{8n_{\mathrm{A}}(\pi+2\Omega t_{\mathrm{eff}}^\pi) -5(3\pi+5 \Omega t_{\mathrm{eff}}^\pi)}{2}\frac{\gamma}{\Omega}\right),
\label{gain_nn}
\end{equation}
where $F_{\mathrm{pro}}$ and $F_{\mathrm{pro}}^\mathrm{nn}$ are the process fidelities for our protocol, and for the protocol relying on a sequence of nearest-neighbor CNOT gates, respectively. Figure~\ref{fig15} shows the results of numerical simulations for the ratio $F_{\mathrm{pro}}/F_{\mathrm{pro}}^\mathrm{nn}$ (dots) as a function of the decay rate in the case of two and three ancilla atoms. The solid lines represent Eq.~\eqref{gain_nn} with $\Omega t_{\mathrm{eff}}^\pi/\pi\approx 0.379$, which was obtained from a fit.
The ratio is always greater than $1$ and increases with the decay rate $\gamma$ and the distance $n_\mathrm{A}$ between control and target qubits.

%%%%%%%%%%%%%%%%%%%%%%%%%%%%%%%%%%%%%%%%%%%%%%%%%%%%%%%%%%

\section{Perspective and experimental considerations}
\label{Perspective}

The estimates for the process fidelity of the protocols presented in this work may not account for all possible dissipation and decoherence channels or experimental imperfections. In this regard, it would be interesting to include in our model a non coding state for both qubit and ancilla atoms in order to account for qubit atom losses due to dissipation. Also, we could consider an intermediary level between the ground-state manifold and the Rydberg state of the qubit atoms as the laser excitation to the Rydberg state is usually a two-stage process. A more rigorous description of the dipole-dipole interaction between atoms leading to the Rydberg blockade could also be considered. However, such simulations are much more demanding in terms of resources.

For simplicity, only square pulses have been considered in this work. From an experimental perspective, it is certainly relevant to investigate the implementation of our protocol using Gaussian pulses or optimized pulses~\cite{The16}, allowing us to further increase the process fidelity. In order to experimentally implement our protocol, one could use the same species for both qubit and ancilla atoms. In such a configuration, the position of the atoms in the different traps or in the lattice will determine its role (coding or non coding) in the protocol. This solution could be implemented with rubidium atoms~\cite{Ise10,Mul14} in dipole traps or using two-dimensional arrays of cesium atoms~\cite{Mal15}. Another possibility is to rely on two different atomic species to implement the qubits and the chain of ancilla non coding atoms. In this case, suitable Rydberg states need to be identified that allow for strong dipole blockade between qubit and ancilla atoms and in between ancilla atoms. A good candidate for the implementation of our protocol is the configuration described in Ref.~\cite{Bet15} in which two optical lattices, one for rubidium and the other for cesium atoms, are considered to perform non demolition state measurements.  

%%%%%%%%%%%%%%%%%%%%%%%%%%%%%%%%%%%%%%%%%%%%%%%%%%%%%%%%%%

\section{conclusion}
\label{Conclusion}

In this paper, we have considered an array of qubits encoded in the ground state manifold of trapped neutral atoms, supplemented by an array of ancilla non-coding atoms. We have proposed a protocol for the implementation of two-qubit entangling gates (CZ, CNOT) between any pair of qubits in the array that relies on the Rydberg excitation hopping along a chain of ancilla non coding atoms in the strong-blockade regime. The hopping of the Rydberg excitation from one atom to its next nearest neighbor is produced by an appropriate pulse sequence that ensures that at most one atom at a time in the entire system is in a Rydberg state. We have solved a master equation for up to nine ancilla atoms in order to evaluate the process fidelity characterizing the performance of our protocols in the presence of dissipation. We have found that the process fidelity is determined by the cumulated time spent by the atoms in the decaying Rydberg states during the execution of the protocols. The design of our protocol ensures that this time scales linearly with the number of ancilla atoms. Moreover, we have shown that our protocols for entangling gates between distant qubits lead to better process fidelities than those based on a sequence of nearest-neighbor two-qubit gates, even when the qubits are separated by a few other atoms. Our protocols could be implemented experimentally for a few  ancilla atoms using state-of-the-art trapping and selective laser-addressing techniques.

{\em Acknowledgments:} Computational resources were provided by the Consortium des Equipements de Calcul Intensif (CECI), funded by the Fonds de la Recherche Scientifique de Belgique (F.R.S.-FNRS) under Grant No. 2.5020.11.


\begin{thebibliography}{99}
\bibitem{footnote4} {Note that this modified CZ gate can be turned into a CNOT gate by single-qubit gates as
\begin{equation*}
U_{\mathrm{CNOT}}=(I\otimes U_H)(U_{S}\otimes U_{S})\tilde{U}_{\mathrm{CZ}}(U_{S}\otimes U_{S})(I\otimes U_H)\nonumber
\end{equation*}
with
\begin{equation*}
U_H=\frac{1}{\sqrt{2}}\begin{pmatrix}1 & 1\\
1 & -1
\end{pmatrix},\quad U_{S}=\begin{pmatrix}1 & 0\\
0 & i
\end{pmatrix}.
\end{equation*}}
\bibitem{footnote2} We added to the interaction potential (\ref{Vdd}) the term $(U_{re}/2^3)\left(|r_\C e_2\rangle\langle r_\C e_2|+|r_\T e_{n_\A-2}\rangle\langle |r_\T e_{n_\A-2}|\right)+(U_{ee}/2^3)\sum_i |e_i e_{i+2}\rangle\langle e_i e_{i+2}|$.
\bibitem{footnote3} Using SWAP gates to transfer the state of the control atom next to the target one would require $6n_{\A}-5$ CNOT gates instead of $4(n_{\A}-1)$ as each SWAP is made of three CNOTs.
\bibitem{footnote1} In the case of a sequence of nearest neighbours gates, as the state of the coding qubits separating the control and target qubits is a priori not known, the process fidelity is computed using the subspace spanned by the coding states of the entire register of qubits.
\end{thebibliography}
\end{document}